# Decentralised Cooperative Collision Avoidance with Reference-Free Model Predictive Control and Desired Versus Planned Trajectories*

Charles E. Wartnaby, Daniele Bellan

*Abstract*— The abilities of connected automated vehicles provide a new opportunity for highly advanced collision avoidance, in which several cars cooperate to reach an optimal overall outcome that no single car acting in isolation could achieve. For example, one car may automatically swerve to allow another to avoid an obstacle. However, this requires solving the challenging real-time problem of deciding what joint trajectories an ad-hoc group of cooperating vehicles should follow, with no obvious leader known in advance. To avoid the complexities of agreeing what plan to follow in an ever-evolving situation, a protocol requiring no leader and no explicit inter-vehicle agreement is desirable, which nevertheless yields cooperative, robust behaviour.

One possible method is demonstrated here, in simulation. This uses the notion of "desired" versus "planned" trajectories, allowing vehicles to influence each other for mutual benefit, without requiring a leader or explicit agreement protocol. Essentially the *desired* trajectory is that which the vehicle would choose if other cooperating vehicles were not present, avoiding the predicted paths of non-cooperating actors. The *planned* trajectory additionally accounts for the planned trajectories of other cooperating vehicles, giving the safest currently available path. Both trajectories are broadcast. As each vehicle attempts to (weakly) avoid the desired trajectories of other vehicles, cooperative behaviour emerges.

A simple form of model predictive control is used by each vehicle to numerically optimise both trajectories. This uses a cost function to penalise predicted collisions, accounting for their severity. There is a weak preference for maintaining the current lane on the road, but no explicit reference trajectory. This decentralised planning and simple optimisation scheme results in robust handling of a wide range of collision scenarios, with no hard limit to the number of cooperating vehicles. The computing cost is linear in the number of vehicles involved.

I. INTRODUCTION

*A. Motivation*

The development of automated vehicles provides capabilities of both object and environment perception on the one hand, and electronic actuator controls on the other. Connected and automated vehicles (CAVs) additionally share information using local radio links, promising various benefits [1]. But bringing together these control and communication functions also gives the opportunity to automate collision avoidance in which multiple cars cooperate to avoid an accident; for example, one car may pre-emptively swerve to allow room for a second car to avoid an obstacle. This cooperation means a group of CAVs working together can avoid accidents that human drivers acting individually could not. This is especially the case in high speed highway situations in which braking is inadequate to avoid an accident, and automated steering intervention is required [2]. High-speed accidents such as this, often involving multiple vehicles, tend to have particularly severe outcomes, and have the additional societal burden of large-scale traffic disruption if highways are brought to a halt. Higher percentages of casualties correlate with increasing numbers of vehicles involved in the accident [3]. There is thus a strong incentive to introduce an automated cooperative collision avoidance system to reduce the incidence and severity of such accidents. This is the focus of the Multi-Car Collision Avoidance (MuCCA) project [4], which should culminate in real-life test track demonstrations of cooperative collision avoidance.

*B. Challenges*

However, it is a challenging algorithmic problem to construct cooperative collision-avoiding trajectories amongst a group of vehicles in real time. Problems include the infinitude of possible scenarios that might be encountered; the difficulty of jointly agreeing trajectories between vehicles, without exposing an "innocent" cooperating vehicle to undue risk; the partial information available to each vehicle in judging the optimal behaviour; the latencies and possible drop-outs in radio messaging; and the lack of any centralised authority to dictate the overall plan, which might suggest the need to choose a single vehicle temporarily as a "leader", requiring some negotiation and agreement protocol. Furthermore as the situation evolves, so might the optimal plan to follow, perhaps requiring a change of leader or agreed plan in mid-manoeuvre. These complexities are at odds with the desire for a robust system that can tackle any arbitrary situation while being resilient to partial or delayed information.

The authors of [5] evaluated different cooperative motion planning algorithms such as tree search, elastic bands, and Mixed-Integer Linear Programming (MILP) in different traffic scenarios. The MILP formulation includes the obstacle avoidance as hard constraints of the optimisation problem, and has been demonstrated in simulation in this project [6]. Such an approach is expected to ensure safety, but it has a high price in terms of computational time due to the binary variables to optimise; it is also constrained to a linear model and cost function [5]. In addition, it might happen that in some scenarios a collision cannot be avoided and thus the problem would yield no numerical solution, yet a real vehicle must be controlled regardless, aiming for the least worst outcome. The problem could be relaxed by including the collision avoidance in the cost function [2]. However, that study acknowledged that a realistic agreement protocol between

*Project supported by Innovate UK.

C. E. Wartnaby and D. Bellan are with Applus IDIADA, Cambridge, CB24 6AZ, UK (e-mail: charlie.wartnaby@idiada.com).

vehicles was still required; in essence the problem was only solved from the perspective of a single actor.

An alternative, leaderless, protocol for achieving cooperative manoeuvres (e.g. lane changes), mediated via a proposed Manoeuvre Cooperation Message, is presented in the TransAID project [7]. There, each vehicle transmits both a planned trajectory (that which it currently intends to follow), and a desired trajectory (that which it would prefer to follow, were it not blocked by other cooperating vehicles). A cooperating vehicle re-plans its trajectory, if possible, to accommodate the desired trajectory of another vehicle, and broadcasts its updated plan. The first vehicle may then update its firm planned trajectory to match its previously broadcast desired trajectory, as it now "knows" that the other vehicle intends to make way for it. At no time is a leader required, and at all times each vehicle plans the safest trajectory available to it, providing robustness to cooperation failure. A similar concept holds in a multi-aircraft cooperative trajectory planning with hard safety guarantees [8], though that uses a safe "loiter" trajectory as a fallback, rather than a preferred trajectory which still makes useful safe progress; also, a processing order among actors must be agreed.

### C. This work

Here, the aim was to assess whether the "desired versus planned" (DVP) trajectory concept could be successfully applied to achieve effective cooperative collision avoidance among an arbitrary number of vehicles, with no leader or central authority. It was not specific to lane changes, but allowed the vehicles full freedom to use the available road, as vehicles would not be constrained to road lanes in real collision avoidance situations.

Collision avoidance was achieved via a "soft" cost function, not via "hard" inviolable constraints. This solved the problem of no numerical solution emerging in scenarios where collision was inevitable. Instead, the least worst outcome was selected, corresponding to the least severe collision.

As the planned trajectory for each vehicle was always the optimum it could follow with the current best knowledge of the trajectories of other vehicles, maximum safety at any time (within the limitations of the model) was assured.

A simple form of Model Predictive Control (MPC) was used to optimise the trajectories, minimising the computed cost function value. This MPC did not use a reference trajectory, but the cost function included terms to (weakly) encourage keeping to the current lane, and maintain speed.

A simulation environment was constructed to develop and test these algorithms in a wide range of scenarios, drawn from project use cases. The results of those simulated tests are presented here.

The rest of this paper is organised as follows: Section II explains the desired versus planned concept using simple results, and the model structure. Section III details the cost function components used. Section IV details computing practicalities and presents additional results. Section V concludes with suggestions for future work.

## II. TRAJECTORY PLANNING

The trajectory planning for each MuCCA Equipped Vehicle (MEV) was performed at each simulated time iteration (typically 40 ms) using a simple formulation of MPC. The novelty here was in the use of "desired" and "planned" trajectories, exchanged continually with other MEVs in a simulation of Vehicle-to-Vehicle (V2V) communications, resulting in spontaneous, cooperative behaviour. Also, the approach here used no explicit reference trajectory, but instead encoded all desired behaviour in the cost function, optimised with a simple non-linear solver allowing complete freedom in algorithm design. As this was intended only as an algorithm demonstration, efficiency was not a concern. The details are provided in the following subsections.

### A. Desired versus planned and first results

The entire optimisation of the trajectory was done twice at each update, for each vehicle: once to determine the *planned* trajectory, and once for the *desired* trajectory. The only difference was that the desired trajectory omitted the calculation of the trajectory overlap and collision cost for the planned trajectories of cooperating vehicles. The desired trajectory thus indicated what the vehicle would "prefer" to do, if only all cooperating vehicles were to make way for it.

When computing the planned trajectory, the cost function strongly preferred to avoid trajectory overlap with the planned trajectories of other MEVs, and weakly preferred to avoid the desired trajectories. This led to cooperative behaviour as explained in the following example, and demonstrating this was the crux of this work. The example uses single-step screenshots from the simulation described below[1] in IV.A, and so doubles as the first experimental results here.

In Fig. 1, two MEVs are labelled M1 and M2. They each have an initial speed of 15 m/s, and are approaching a stationary human-driven vehicle (obstacle) in the same lane as M1, 22 m ahead. Their *planned* and *desired* trajectory points over a prediction horizon of 23 points with 0.07 s separation are shown as circles and crosses respectively.

At step 0, neither vehicle has yet sensed the obstacle. They plan to continue at constant speed without deviation.

At step 1, both MEVs have replanned their trajectories based on step 0 information. But as M2 has no reason to do otherwise, it continues to plan a straight course. M1 however now *desires* to swerve around the obstacle it sensed last time, allowing it to avoid it without incurring the penalty of harsh braking. But this course would be in conflict with the *planned* trajectory of M2 it received at step 0. So to be safe, it *plans* a braking trajectory, attempting to stop before the obstacle. Its desired trajectory is to swerve; but its planned trajectory is to brake.

At step 2, the trajectories are replanned based on step 1 information. M2's preference to remain in lane is weak, so it obligingly avoids the desired trajectory broadcast by M1 last time. But for M1, nothing has qualitatively changed since it is still working with step 1 information; it still *desires* the swerve, but *plans* to brake.

Finally at step 3, M2 continues to plan a swerve, making way for M1; and M1, having received that swerving *planned*

---

[1] For this introductory section only, the collision cost weights (Section III.B) were temporarily zeroed, to prevent an MEV preferring to "gently" side-swipe its neighbour over colliding with the obstacle regardless of cooperation, so that the DVP principle would be exhibited clearly.

trajectory from M2 at step 2, may now firmly plan to swerve around the obstacle. The end result is that both vehicles swerve in a cooperative, swarm-like manner; yet neither vehicle was elected leader, and there was no explicit shared planning. They acted as equal peers, running the same single-pass algorithm at each step. There was no voting, no evaluation of rival plans generated by different vehicles, and no question of switching between discrete plans. The scheme is simple and robust.

Computationally, each MEV is iterating over the trajectories of every other MEV with which it is cooperating; the cost is linear in the number of MEVs, $O(n)$ for each MEV. As this simulation is all run on one CPU however, the cost increases as $O(n^2)$. Solving took ~100 ms per vehicle considered on an Intel Core i7 processor.

Note that at step 2, M1's *desired* trajectory is also to swerve. The formulation here is that desired trajectories avoid overlap with other *desired* trajectories, but ignore other *planned* trajectories. The reason for this is that if a third vehicle M3 were stacked alongside M1 and M2, but now on a 4-lane highway, the deviation of M2's desired trajectory would in turn cause M3 to also swerve, achieving cooperation between all three vehicles. Fig. 2 shows this new scenario after 6 iterations, when a cooperative swerve by all three vehicles becomes fully established. If desired trajectories were not accounted for, this would not occur: instead, M2 would remain "boxed in" by M3's straight planned trajectory, and would not express its desire to swerve into M3's lane.

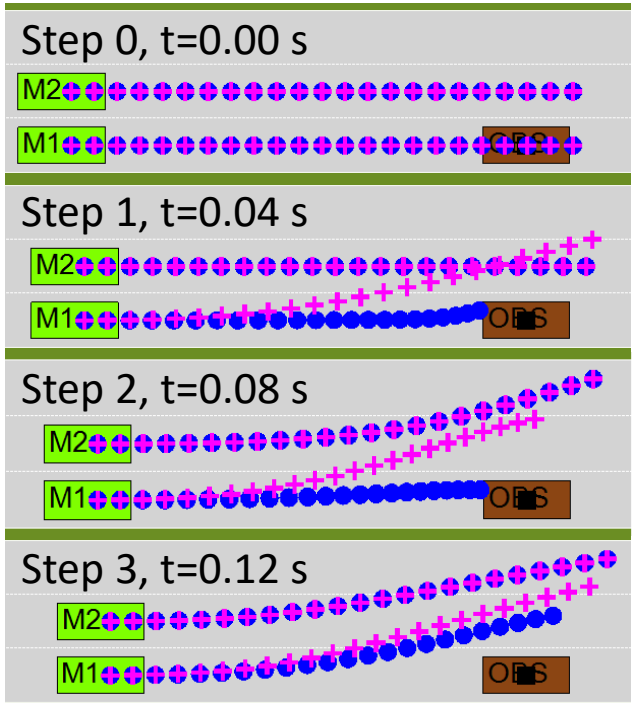

Figure 1. Desired and planned trajectories resulting in cooperation; circles are planned trajectory points, crosses are desired trajectory. X-axis extends right, Y-axis up; X and Y directions are shown at equal scale, with a lane spacing of 2.5 m and vehicle length of 4.0 m.

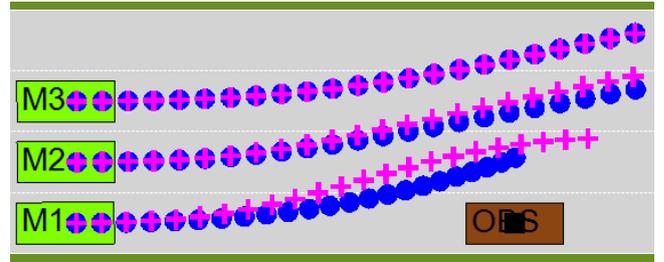

Figure 2. Spontaneous cooperation amongst three vehicles

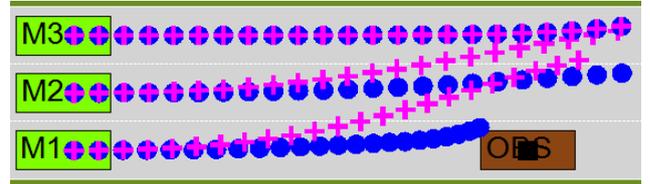

Figure 3. Safe default behaviour (braking) by M1 if cooperative swerving is not possible

If we keep M3 but return to a 3-lane highway (Fig. 3), M3 can no longer swerve significantly because it is next to the road boundary, and so M2 in turn is prevented from swerving much. Thus M1 cannot plan to swerve round the obstacle and so pursues the safest remaining option, i.e. braking hard; the default behaviour is the safe behaviour, should cooperation fail. M2 and M3 can achieve only a modest deflection to aid M1 in this case, making use of "spare" road space (as the vehicles are narrower than the lanes).

### B. Simple Model Predictive Control

#### 1) Vehicle Model

The simulation work here was for proof of principle only, to explore the qualitative behaviour of the desired-versus-planned (DVP) approach. Therefore no attempt was made in this case to implement a high-fidelity vehicle model. In practice, the work here might be used to compute an idealised reference trajectory, with a lower-level tracking controller being used to achieve that trajectory, accounting for vehicle dynamics and expected latencies.

Instead, a simple kinematic model was assumed. Each vehicle was represented by a centre point position with orientation, velocity and yaw rate. Control outputs were steering (yaw rate of change, lateral) and acceleration/braking (acceleration rate of change, longitudinal). However, collision overlaps did consider the length and width of the vehicle.

In the event of collision, both vehicles involved were instantaneously given zero velocity – no attempt was made to model post-collision physics. This meant that if a scenario started with one HDV catching up another, they would both halt and become an "unexpected" stationary obstacle.

The numerical solver was a very simple "Twiddle" algorithm [9] which made experimental step changes to each variable in turn, before a full re-evaluation of the cost function; the change increment was then amplified, or reversed and diminished, depending on whether the result was improved. When the changes for all variables was below a threshold, the solution was assumed to have converged. While inefficient, this method does not require that the cost function is differentiable or even continuous, allowing complete freedom in algorithm design.

As we expect collision avoidance to often involve steering either to the left or the right, the solver was given a

"hint" by starting it off with all steering changes to the left; after optimisation was complete, the solver was run again, but this time with initial changes biased to the right. The lower-cost solution was then retained. This helped the system avoid unwanted local minima, e.g. as might be found squeezing through a gap between an obstacle and the road margin by steering to the right, when open road was available to the left. Running the solver twice like this improved robustness, at the obvious expense of efficiency.

### C. No reference trajectory

Traditionally, MPC is used to plan what control outputs should be set to minimise the difference between *actual* system state and some *reference* desired state; for a trajectory, that reference would be a specific position in time computed for each future iteration of the MPC solver, or a future trajectory of specific coordinates to be achieved at each timestep. The reference trajectory might be defined in terms of some sample points to interpolate between, or a smooth analytical function, e.g. a 5th order polynomial to minimise longitudinal and lateral jerk. Having a reference trajectory implies some specific plan for the vehicle to follow, e.g. a route on a map combined with desired lane choices.

Here however, there was no reference trajectory at all. Instead, a term in the cost function (see III.G below) weakly preferred trajectories that minimised distance from the centre of whichever the nearest road lane was. However, there was no requirement to aim for a particular target lane. This system is intended only as a temporary override to normal driving; the driver, or normal automated driving function, would be expected to resume control post collision avoidance, and then choose the lane they consider appropriate. Requiring the system to perform lane changes after any avoidance event would have increased complexity and possibly led to unwanted unexpected outcomes.

### D. Sparse control updates

The sample time used for prediction was kept reasonably small (0.07 s), to ensure that the cost function would faithfully reflect close approaches (or collisions) between vehicles, and for numerical stability. But having the control output changes at every such timestep would mean a very large number of independent variables to optimise in the numerical solver, much increasing CPU time and reducing robustness.

In practice it was found unnecessary to allow the control outputs to vary on the same fine-grained timescale however. Just as a human driver might vary the general emphasis of steering or braking only on a coarse timescale during a manoeuvre, so the simulation needed only to vary those outputs at much lower frequency than used for prediction. A decimator was therefore used to hold the control change variables for a number of prediction steps. This divided down the number of free variables. For example, for a 2 seconds prediction horizon with a timestep of 0.1 s, there would be 40 free variables to optimise if the acceleration and steering outputs were allowed to change at every step; with a decimator of 10, they would be altered only every 1 second, with just 4 free variables to optimise. A decimator value of 8 was found to be a good compromise. This much reduced CPU time.

### E. Simulated vehicle-to-vehicle communications

At each update, the current desired and planned trajectories of each cooperating vehicle were cached by every other vehicle, and then not used until the next update. Hence the simulation incorporated the concept of a transport delay (of exactly one simulation timestep) involved in communications, as would be present in a real network of cooperating cars. It was as if those trajectories had been sent over a radio link at some arbitrary time between update events. Although the simulation effectively updated all vehicles simultaneously, this synchronisation would not be required in reality, and may even lead to unwanted behavioural artefacts, e.g. alternation between two possible trajectories as plans are exchanged between vehicles, which has been seen in this work.

In the real MuCCA project, actual V2V messages are exchanged using 802.11p and modified ETSI-G5 protocol.

### F. Human-driven vehicles

Other vehicles in the simulation could either be cooperative, or not. Non-cooperating vehicles represented human-driven cars, or obstacles (if stationary).

The state of human-driven vehicles (HDVs) was updated at each simulation iteration according to a simple kinematic model, preserving their initial speed, orientation, acceleration and yaw rate. In practice only trajectories parallel to the road were used, with zero acceleration and yaw rate. Often HDVs were stationary, representing obstacles.

As far as each MEV was concerned, the state of HDV was evolved forwards in time to build up a predicted trajectory, which could then be fed into the cost function alongside the predicted trajectories of other MEVs. The prediction was based on the state of the vehicle at the previous timestep, simulating the concept of some delay in sensing and perception.

Note that in the MuCCA project proper, a machine learning model has been developed to predict HDV behaviour, based on data acquired for this purpose from volunteers "driving" in a simulation environment.

## III. COST FUNCTION COMPONENTS

The cost function used was an arithmetic sum of the components described in the following subsections. As a simple brute-force solver was used, there was no requirement for the cost function terms to take any particular functional form (e.g. quadratic), or to be differentiable.

Future work may find (by sensitivity analysis) that some terms could be thrifted out of the calculation of the desired and/or planned trajectory, or computed with reduced prediction fidelity, to reduce computation. For the purposes of this work however, all the components were used, and computed with the same timestep over the trajectory to be optimised.

### A. Trajectory overlap

Trajectory overlaps of three types were computed: with non-cooperating human-driven vehicles (HDVs), with the planned trajectories of other MEVs, and with the desired trajectories of other MEVs. These all used the same formulation, but with different weights *W*, as detailed below.

For each pair of trajectories under consideration, the overlap cost was computed initially as a sum of the reciprocals of the Cartesian separation squared of the two vehicles *(A,B)* across all time-matched trajectory point positions $p_i$:

$$C_{overlap}(A, B) = W_{overlap} \sum_{i=0}^{n} \frac{1}{(p_i(A) - p_i(B))^2} \quad (1)$$

However, it was found that longer prediction horizons tended to dilute the effect of important close vehicle approaches. As an empirical trick, the maximum value of the reciprocal was thus held for the remaining iterations of the sum.

Different values of the weight $W_{overlap}$ were used for overlap with MEV planned, MEV desired, and non-cooperating vehicle trajectories.

A minimum 'epsilon' distance was used to avoid division by zero. The cost was also made zero if the Cartesian distance (between vehicle centres) or lateral distance (across lanes) exceeded thresholds. This allowed fairly close side-by-side passes to go unpenalised, and saved some computation time.

### B. Collision

A specific cost was added if two vehicles were found to overlap at any of the future predicted trajectory points, at matched time. This considered the orientation of each vehicle, and its width and length; the latter were "inflated" by a small tolerance distance to discourage collision avoidance by very small margins when planning trajectories, but not when scoring an overall scenario outcome. The cost was added only once if the two vehicles were found to overlap at any point in the future.

Each pair of trajectories was considered independently (so it was possible to accumulate cost from two different collisions, while in reality a first collision may well preclude a second). However, when later judging the overall cost of a scenario, only the first collision detected was scored.

The collision cost was approximated by a measure of the kinetic energy dissipated, plus some contribution from the absolute velocity of the slower vehicle, $v_{min}$. This meant that gentle "side-swipe" collisions would have relatively low cost compared to a head-on collision with a stationary obstacle:

$$C_{collision}(A, B) = W_{collision} \left\{ |v_A - v_B|^2 + \frac{v_{min}^2}{4} \right\} \quad (2)$$

There is an opportunity at this point to encode ethical rules, by penalising less "allowable" collisions more harshly. In this case a somewhat lower weight (and hence cost) was used for collision between cooperating MEVs than with independent HDVs, but no attempt was made at this juncture to justify or implement specific ethical rules.

### C. Desired avoidance importance

The *importance* of avoiding overlap or collision only with desired trajectories was considered, as justified here.

Consider an MEV approaching an obstacle at high speed: its planned trajectory may be to brake hard, but nevertheless suffer a severe collision; its desired trajectory may be to harmlessly swerve, but that manoeuvre may be blocked unless another MEV makes room. The difference in cost between its planned and desired trajectories will be very great in this situation. Conversely, a second MEV may find its optimal planned trajectory is to swerve to one side of an obstacle, but its desired trajectory to swerve the other way, with little difference in cost. Should *both* of those MEV's desired trajectories be under consideration by a *third* MEV, then it seems obvious that it should give far greater weight to the first vehicle, which has a great need for cooperation, and little weight to the second vehicle that has little need for help.

Hence the desired trajectory overlap and collision cost weightings were scaled down by a simple multiplicative factor "broadcast" by the originating vehicle at the last iteration, reflecting the priority with which its desired trajectory should be respected, based simply on the ratio of the total cost function output for its desired and planned trajectories:

$$K_{desired-importance} = 1 - \frac{C_{total}(desired)}{C_{total}(planned)} \quad (3)$$

### D. Acceleration change (jerk) and forwards acceleration

A cost was added proportional to the squared magnitude of the rate-of-acceleration control output $c_a$ at each prediction point (noting that this was held from the most recent *control* point):

$$C_{long,jerk} = W_{long,jerk} \sum_{i=0}^{n} c_{ai}^2 \quad (4)$$

Additionally, acceleration in the direction of travel was penalised (i.e. "accelerating out of trouble"); if the acceleration and travel were in opposite senses (e.g. braking when travelling forwards), this cost was zero, otherwise:

$$C_{forward,accel} = W_{forward,accel} \sum_{i=0}^{n} a_i v_i \quad (5)$$

### E. Steering change (jerk)

Steering control output changes $c_s$ (lateral jerk) were handled similarly to acceleration changes:

$$C_{lat,jerk} = W_{lat,jerk} \sum_{i=0}^{n} c_{si}^2 \quad (6)$$

### F. Maintaining progress

We preferred to have vehicles maintain speed if possible, rather than braking unnecessarily to a halt. A cost factor was therefore added that penalised a decreasing rate of change of the x-coordinate, i.e. negative acceleration in the "continue down the road" direction. Positive acceleration was not penalised by this term.

### G. Lane-keeping

It is the intention of this system to demonstrate collision avoidance during brief interventions, rather than maintained autonomous driving. However, a weak cost was added proportional to the squared distance from the nearest lane centre, to give reasonable pre- and post-avoidance behaviour. Note that no preference was given to any particular lane; there was no incentive for a vehicle to return to its original lane after avoiding a collision.

### H. Road-keeping

A strong cost was added proportional to the distance squared between the vehicle and the most extreme y-coordinate it could occupy *that would cause it to just touch the road boundary*. This was added only if the car was off the left or right sides of the road, in effect. This did *not* consider vehicle orientation, only width, so minor transgressions of the front or back of the vehicle over the road boundary could occur due to rotation from the straight-ahead orientation.

### I. Constraint Violations

A large fixed cost was added if a proposed trajectory violated any constraint, where those were: a modest (2 m/s²) forwards acceleration; hard but achievable (10 m/s²) braking deceleration; minimum forwards velocity (0 m/s); and a maximum yaw rate (5 rad/s). Strictly these were not hard constraints, but the high cost associated with violating them prevented the optimiser from selecting trajectories that did so.

IV. SIMULATION EXPERIMENTS

A. Computing Environment

The MuCCA project uses the PreScan simulation environment with real control hardware in the loop. However, for the purposes of this study, a simplified bespoke simulation was written as a Windows application in C#. Computational efficiency was not optimised, although some aspects (e.g. the control update decimation) were tuned to improve performance. On-screen controls allowed rapid experiments in which either predefined use-cases were selected, or manual changes were made to the initial states of vehicles. User controls allowed many vehicles to be selected in either cooperative or non-cooperative mode; the dynamic memory allocation available in C# meant that no fixed-size data structures were required for a specific number of vehicles, an advantage over some more traditional simulation languages. The road length, lane width and number of lanes were all variable, and the cost function weightings accessible for editing. Overall, this environment was conducive to rapid experimentation with different scenarios and algorithm changes.

B. Meta-Optimisation

The relatively complex cost function presented in Section III gave a non-trivial number of weights to adjust. Reasonable behaviour was found via *ad hoc* experiment, but the means to numerically optimise those weights was explored. Although this was not successful, it provides a basis for future work to build upon.

*1) Scenario selection*

After each pass through the parameters, a new scenario was presented, to ensure that the model did not overfit the weights to a single scenario; this was a form of stochastic gradient descent (SGD).

The new scenario was either drawn from one of the predefined project use-cases, or a random scenario populated by up to 6 vehicles, with between 1 and 6 MEVs. Scenarios which resulted in almost no cost overall at the first attempt were skipped over, as they would not result in useful learning of cost function weights. Only non-trivial scenarios were therefore iterated for all of the cost function weights.

*2) Evaluation of Success*

A simple scoring system was used to evaluate the success of the system in bringing a scenario to a satisfactory conclusion; without this, there was nothing quantitative to optimise automatically. Firstly, the collision cost (see Section III.B) was summed for all involved MEVs. This term had a strong waiting. Secondly, a *halting cost* was added, weakly penalising outcomes in which the MEVs lost their original speed (here considered as a scalar). This preferred outcomes which showed continuity of driving rather than arbitrary braking or stopping, iterating over all m MEVs:

$$C_{halting} = W_{halting} \sum_{k=0}^{m} \{v_{final}(k) - v_{initial}(k)\}^2 \quad (7)$$

These costs were computed at a fixed time interval after any collision avoidance behaviour of interest was likely to have completed.

*3) Optimisation Algorithm and Outcome*

The simple Twiddle algorithm was again tried, but this time the variables were the weight factors rather than the MPC control outputs.

In practice, the weights were not found to converge satisfactorily. Random scenarios, in particular, tended only rarely to present opportunities for cooperative behaviour to achieve a lower cost, so there was little "incentive" for convergence; finding no benefit in changes, the weights tended to converge at their current values. Even predefined use cases could exhibit dramatic changes in outcome (e.g. collision versus no collision, or side-swipe of one vehicle versus straight-on collision with an obstacle) encouraging a change in a weight factor for one scenario that would likely prove disadvantageous overall; the cost space was not smooth enough for the meta-optimization to work with this algorithm.

A genetic algorithm [10] may be better suited for a problem of this sort, where parameter changes result in branches into qualitatively different behaviour, giving a multimodal (cost versus parameter) solution space.

C. Use-Case Examples and Problems

The effect of the DVP approach using the cost function components described in III above was seen already in II.A above by way of explanation of the principle, demonstrating cooperation between 2 or 3 MEVs with one obstruction. Here the experimental results for emergent cooperative behaviour in other use-cases of interest is shown: one successful, one not. Space limitations preclude the results of further use cases being presented here, but in fact the same simulation with unchanged algorithms and parameters was demonstrated with a variety of use cases with different numbers of vehicles.

In this work, the prediction horizon for the vehicles was kept rather short, and the cost function weights tuned to encourage swerving over braking where possible, in order to explore the cooperative behaviour that could emerge. Note that with tuning suitable for production, these use cases might instead result in strong braking, initiated at greater range, and little or no steering. It is the qualitative behaviour which is of interest here however; that useful cooperation can emerge at all using the leaderless DVP protocol.

*1) Five MEVs Plus Obstruction*

The use-case shown in Fig. 4 is one of the most challenging devised early in the MuCCA project. A set of 5 MEVs, all initially travelling at 8 m/s, all cooperated to avoid an obstacle. M1 and M2 were not obstructed and continued with only slightly perturbed trajectories. M3 is close to the obstacle and was forced to brake hard to avoid M2, which could not cooperatively swerve because of M1. However, M3 could also steer to some extent, its desired trajectory falling in behind M2, nudging M4 to cooperatively re-plan its trajectory to move lane to make space for it. M4's broadcast plan to swerve also allows M5 to steer as well as brake to avoid M3. Overall a satisfactory outcome was achieved in which no collision occurred, three vehicles continued down the road (with one changing lane), and two vehicles used hard braking and steering, making use of the space vacated by the lane-changer. Note as before that no prior knowledge of this use-case was baked into the simulation; all the MEVs acted as equal peers, following the same algorithms.

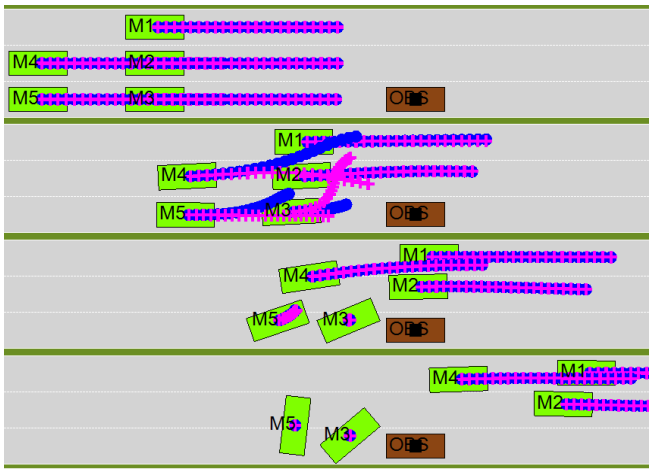

Figure 4. Five MEVs cooperatively avoid obstruction

*2) Deliberate Race Condition, Oscillatory Behaviour*

In this case (Fig. 5), two MEVs were positioned in lanes 1 and 3 with identical initial speeds of 15 m/s such that they would each encounter an obstruction in their respective lane. A race condition thus existed, in which the optimal behaviour would be for one MEV to pass through the gap between obstructions and for the other to follow it, but with no algorithmic reason to choose one or the other. In this case, unwanted oscillatory behaviour emerges, due to the current strict exchange of trajectories between updates.

As step 0, both MEVs planned to brake before their obstructions, because in a previous step they *both* shared plans to aim for the gap, and are *both* now avoiding each other's planned trajectories. Their desired trajectories however are to swerve through the gap. At step 1, each MEV then "knew" that the other intended to brake, apparently leaving the gap free, so they *both* updated plans to aim for the gap. At step 2, this situation was reversed again, and so-on.

If a 5 metre difference in the initial longitudinal offsets of the MEVs was introduced, the lead vehicle swerved and the following vehicle braked and swerved, and they passed through the gap in turn, as we would intuitively expect. In real life, we would not expect a perfect symmetry. However, the oscillatory behaviour can persist even with slight symmetry breaking, and so must be addressed in future.

Introducing some stochastic noise (to break symmetry), and also some "stickiness" (memory) of the trajectories selected at the previous timestep would be the first approach. Then one MEV would win slightly (through noise), and its advantage would persist (by memory), and ideally grow to a satisfactory outcome. Some priority based on lateral offset (i.e. lane number) based on road law could also be introduced, e.g. by tuning the cost function weights to encourage braking more in the legal "slow" lane. However, that would not preclude race conditions emerging where e.g. the "benefit" of being in the "fast" lane but also starting at lower speed just happen to balance out in the cost function.

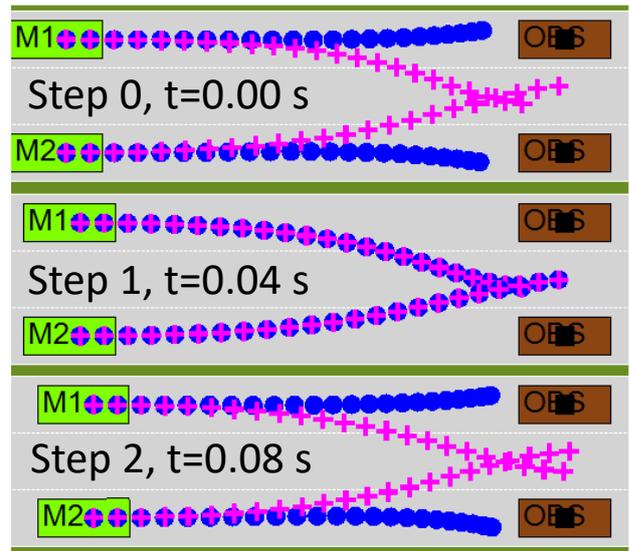

Figure 5. Unwanted oscillatory behaviour emerging in symmetric situation

## V. CONCLUSION

The DVP approach, using a simple MPC with no reference trajectory and no hard constraints, was able to produce cooperative behaviour among a group of vehicles, as shown in a basic simulation. This simple approach promises robustness to unknown or unexpected situations as each vehicle plans always the safest trajectory it can find based on current knowledge, with no need to compare alternative joint plans or invoke a leader or outside authority.

Further work is required to eliminate unwanted oscillatory solutions that can emerge in symmetric situations, to simplify and automatically tune cost function weights, and to bring this solution to the real project fleet – within available computational resources, with real sensing, V2V messaging and actuator controls, and modelled with sufficient fidelity for predictable safe outcomes.

### ACKNOWLEDGMENT

IDIADA would like to thank the MuCCA consortium partners (Cranfield University, Transport Systems Catapult, Cosworth Electronics, Westfield Sports Cars and SBD Automotive) for their cooperation on this project, and especially to Ícaro Bezerra Viana for reviewing this paper.

### REFERENCES

[1] D.J. Fagnant and K. Kockelman, "Preparing a nation for autonomous vehicles: opportunities, barriers and policy recommendations," *Transportation Research Part A: Policy and Practice*, vol. 77, pp 167-181, 2015.
[2] J-B. Tomas-Gabarron, E. Esteban Egea-Lopez and J. Garcia-Haro, "Vehicular trajectory optimization for cooperative collision avoidance at high speeds," *IEEE Transactions on Intelligent Transportation Systems* vol 14.4, pp 1930-1941, 2013.
[3] UK STATS19 database, 2016.
[4] *MuCCA Project* [Online]. Available: https://mucca-project.co.uk/
[5] C. Frese and J. Beyerer. "A comparison of motion planning algorithms for cooperative collision avoidance of multiple cognitive automobiles," *Intelligent Vehicles Symposium (IV)*, IEEE, 2011.
[6] Í. B. Viana and N. Aouf, "N. Distributed Cooperative Path-Planning for Autonomous Vehicles Integrating Human Driver Trajectories," in *9th IEEE International Conference on Intelligent Systems*, Funchal, Madeira, 2018.

[7]  M. Rondinone and A. Correa, "TransAid Project D5.1 Definition of V2X message sets" [Online], Available: www.transaid.eu

[8]  T. Schouwenaars, J. How and E. Feron, "Decentralized cooperative trajectory planning of multiple aircraft with hard safety guarantees," in *AIAA Guidance, Navigation, and Control Conference and Exhibit*, 2004.

[9]  The Twiddle algorithm is attributed here to Sebastian Thrun, in lectures through the Udacity programme, [Online], Available: https://www.youtube.com/watch?v=2uQ2BSzDvXs

[10] B.L. Miller and M.J. Shaw, "Genetic Algorithms with Dynamic Niche Sharing for Multimodal Function Optimization," in *International Conference on Evolutionary Computation*, 1996.